\documentclass{mem}
\usepackage{txfonts}
\usepackage{balance}
\usepackage{graphicx}

\usepackage{natbib}
\usepackage[a4paper]{hyperref}
\begin{document}
\def\teff{$T\rm_{eff }$}
\def\kms{$\mathrm {km s}^{-1}$}
\newcommand{\enzo}{\it ENZO}
\newcommand{\na}{New Astronomy}
\title{Synchrotron emission and neutral hydrogen in the simulated cosmic web}

   \subtitle{}

\author{
F. \,Vazza\inst{1,2,3} 
\and S.\, Banfi\inst{1,2}
\and C. \, Gheller\inst{2}
          }

  \offprints{franco.vazza2@unibo.it}

\institute{
Dipartimento di Fisica e Astronomia, Universita di Bologna, Via Gobetti 93/2, 40122, Bologna, Italy 
\and 
INAF, Istituto di Radioastronomia di Bologna, via Gobetti 101, I-41029 Bologna, Italy 
\and Hamburger Sternwarte, Gojenbergsweg 112, 21029 Hamburg, Germany }

\authorrunning{Vazza, Banfi, Gheller}

\titlerunning{Synchrotron and HI in the cosmic web}

\abstract{We present the first results of a campaign of ENZO cosmological simulations targeting the shocked and the neutral parts of the cosmic web, obtained with Supercomputing facilities provided by the INAF-CINECA agreement.

}
\maketitle{}

\section{Introduction}

Detecting the missing baryons of the cosmic web  \citep[e.g.][]{2016xnnd.confE..27N} o and understanding the origin of cosmic magnetism \citep[e.g.][]{sub16} are two pivotal goals for modern astrophysics. Detecting the radio signature of the cosmic web is also strictly related to the possibility of constraining the origin of cosmic magnetism \citep[e.g.][]{va17cqg,gv20}, with potential impacts on, both, galaxy formation theory and cosmology. 

{\bf As a part of  a $\sim 10$ years long campaign of simulations of the cosmic web and of its non-thermal energy components, largely supported  by resources made available within the
INAF-CINECA Memorandum of Understanding, we produced 
a new  set of simulations run on the Marconi and M100 Superclusters (dubbed "ROGER"  Resimulations Of the Gigaparesc-Extended Radio web).}

In these models we also followed the formation of neutral Hydrogen (HI) outside of halos and in filaments, in order to use it as a signpost for missing baryons, and to cross-correlate the HI signal with the synchrotron from the shocked gas in the cosmic web.  The hyper-fine structure line of HI line at 21 cm may potentially by a reliable marker of the cosmic web: while the Universe is mostly ionized at $z \leq 6$, the SKA is expected to have an enough sensitivity to detect the weak HI signal from the residual cold neutral hydrogen in filaments   \citep[][]{2015arXiv150101077P}, even if recent stacking experiments have set an upper limit of $\leq 7.5 ~\mu K$ on the average HI temperature from filaments connecting  halos \citep[][]{2019MNRAS.489..385T}.
However, both HI and  synchrotron radio emission are predicted to depend on a number of physical assumptions on the origin of cosmic magnetic fields, and on the detail of cooling and feedback of baryons in large-scale structures, which makes it necessary to resort to numerical simulations in order to predict the amplitude of detectable signals in a quantitative way \citep[][]{2017PASJ...69...73H}. 

\begin{figure*}[t!]
\begin{center}
    
    \includegraphics[width=0.49\linewidth]{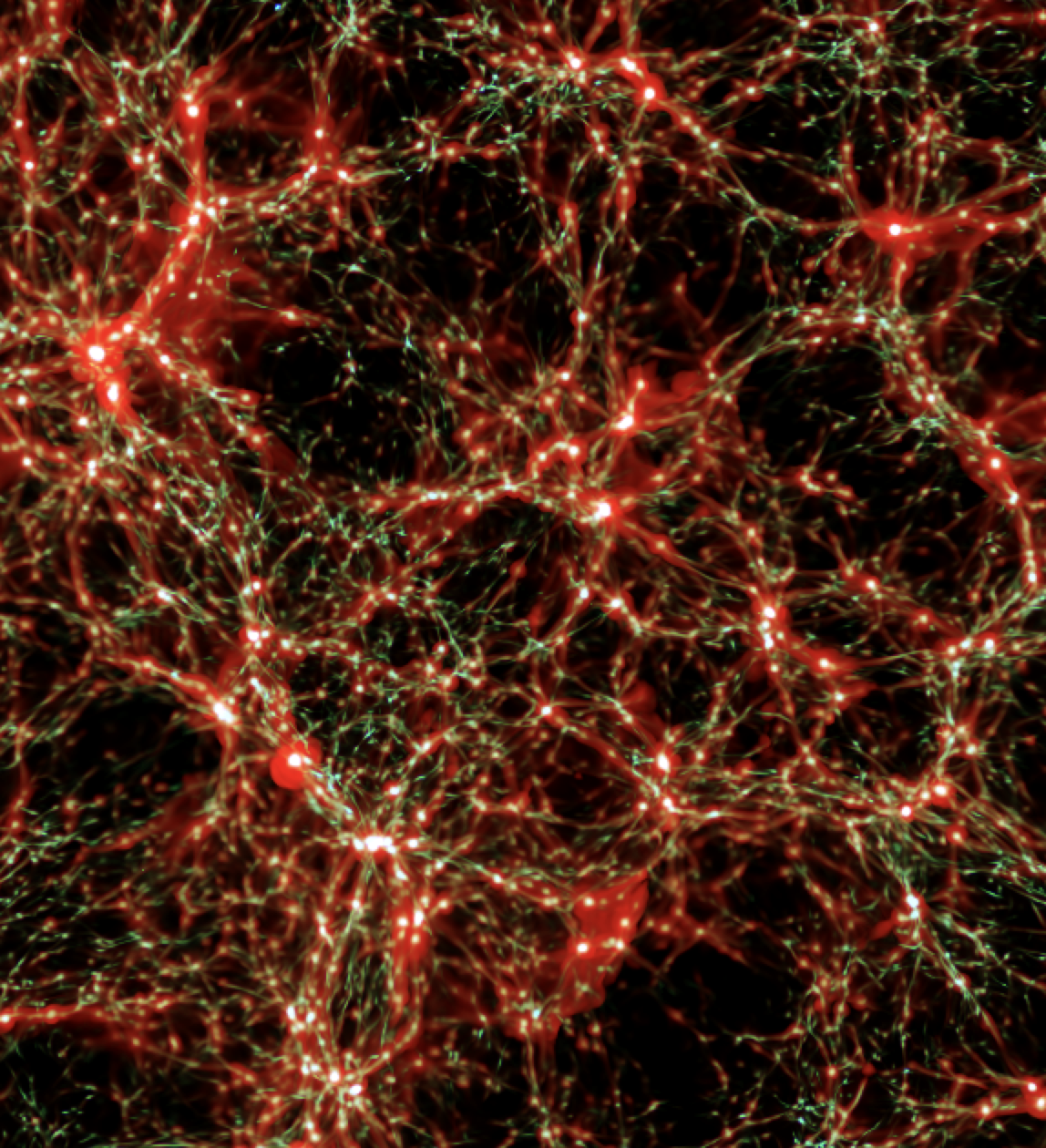}
     \includegraphics[width=0.49\linewidth]{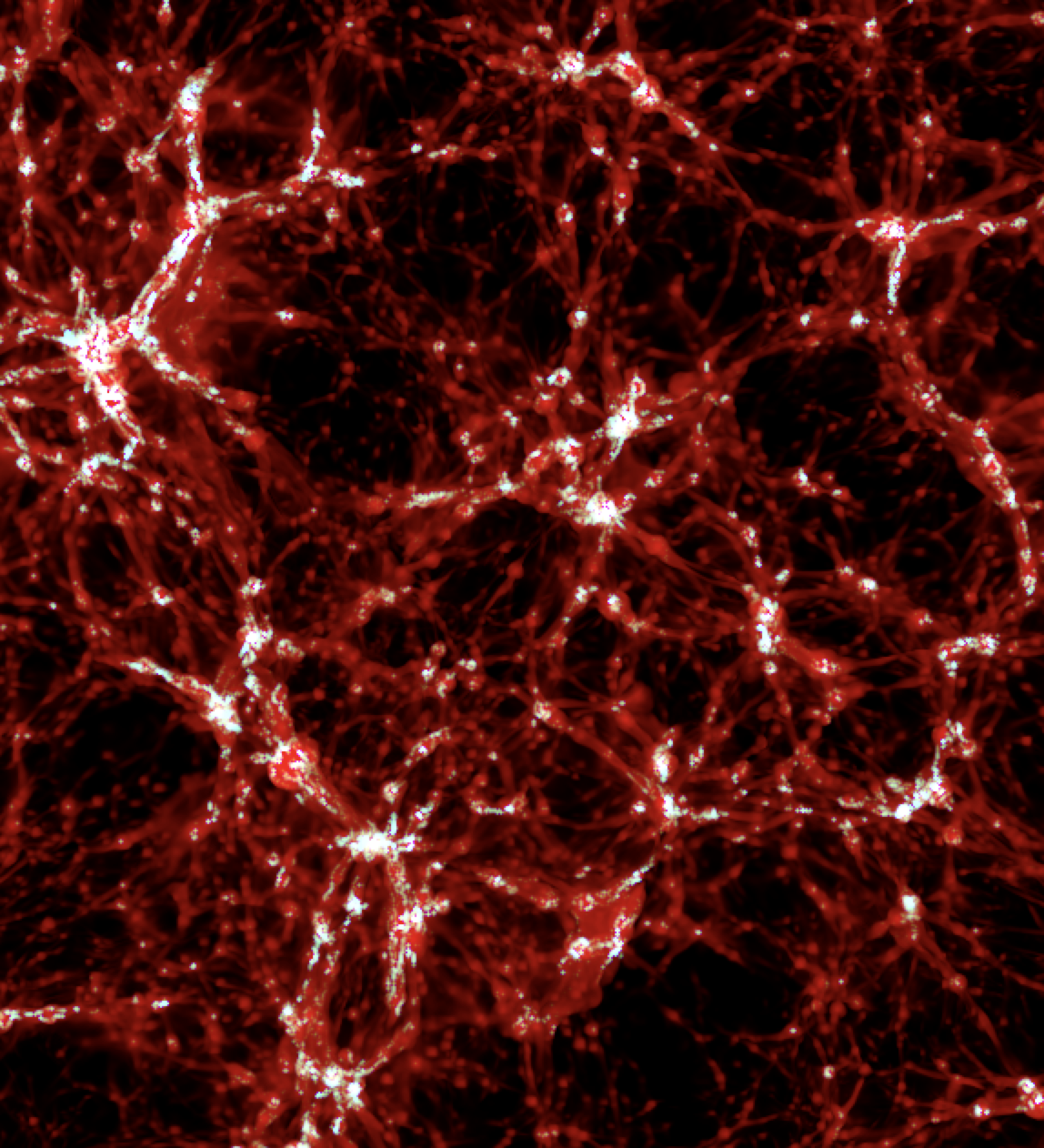}
    
\caption{\footnotesize Left: volume rendering of the distribution of gas temperature (red) and HI temperature (green/blue) for a simulated $100^3 \rm Mpc^3$ at $z=0.3$. Right: same rendering of gas temperature, combined with synchrotron radio emission at 200 MHz (yellow).}
\label{eta}
\end{center}
\end{figure*}
\section{Simulations}

We used the cosmological magnetohydrodynamical grid code {\enzo} \citep[][]{enzo14} to simulate $400^3$, $200^3$, $100^3$, $50^3$ and $25^3$ $\rm Mpc^3$ cosmic volumes, always using a static $1024^3$ grid and in the concordance $\Lambda$CDM cosmology. In most runs we contrasted 
 non-radiative (``NR'') runs with a primordial uniform volume-filling comoving magnetic field $B_0=0.1$ nG, initialised at the  beginning of the simulation ($z=40$) with radiative run (``COOL") also including equilibrium gas cooling, and a sub-grid modelling of thermal AGN feedback, as detailed in \citet{va17cqg}. 
 Using modules included in {\enzo}, we  tracked at run-time  H, H+, He, He+, He++, and electrons, and their  evolution is computed by solving the rate equations with one Jacobi iteration with implicit Eulerian time discretization, with a coupling between thermal and chemical states at subcylces in the hydrodynamical timestep to evolve a number of processes (atomic line excitation, recombination, collisional excitation, free-free transitions, Compton scattering of the cosmic microwave background and photoionization from metagalactic UV backgrounds).
 
{\bf Our runs at CINECA used a total of $\sim 3$ million KNL hours on Marconi  and $\sim 2$ million KNL hours on Marconi100 \footnote{ \url{https://www.hpc.CINECA.it/hardware/marconi}}, typically using 512 cores on 64 nodes for each job. In the latter case, we took advantage of the GPU implementation of the Dedner MHD solver in {\enzo} \citep[][]{wang10}, which makes cosmological static grid simulations faster by a factor $\sim 4 \times$.}

In post-processing, we computed synchrotron radio emission from shocked relativistic electrons and the  HI signal as in \citet[][]{gv20}; 
a) the  spin temperature of HI is computed assuming  the excitation and de-excitation by the CMB photons, the collisions with electrons and other atoms and the interactions with
background Lyman-$\alpha$ photons
\citep[][]{2017PASJ...69...73H}; the HI abundance is computed in a self-consistent way by {\enzo} chemistry modules. b) The emission from relativistic electrons  accelerated by diffusive shock acceleration is the convolution of the several power-law energy distributions of electrons in the synchrotron cooling region, as in \citet{hb07} and in our previous works.

\begin{figure*}[t!]
    \includegraphics[width=0.495\linewidth]{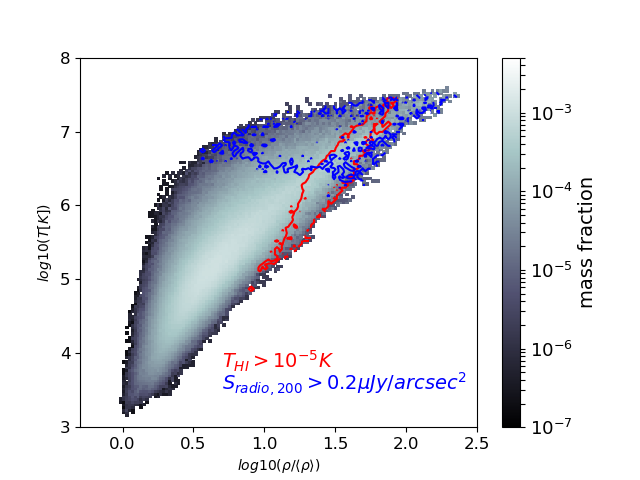}
    \includegraphics[width=0.495\linewidth]{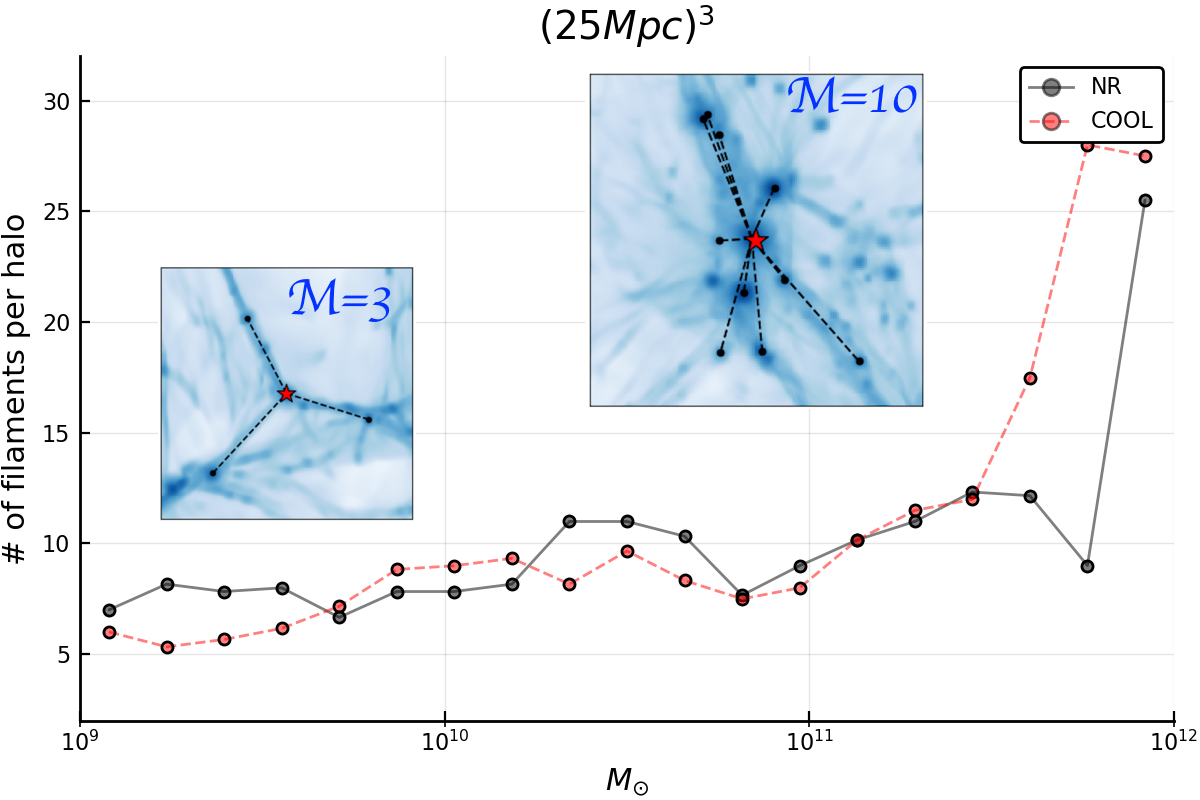} 
\caption{\footnotesize Left: phase diagram for the {\it projected} mean mass-weighted gas density and gas temperature for a synthetic sky model of a simulated volume at  $z=0.5$. The contours enclose the region where $\geq 10\%$ of baryons can be detected either via their synchrotron radio emission at $200 \rm ~MHz$ with SKA-LOW (blue) or through their  HI emission at 1.4 GHz with SKA-MID (red). Right: average number of filaments connected to halos, as a function of halo mass, for our $25^3 ~\rm Mpc^3$ simulated box, using either a sinmple non-radiative run (NR) or including radiative gas cooling (COOL). Adapted from Banfi et al., submitted}
\label{eta}
\end{figure*}
\section{The HI and synchrotron network of the cosmic web}

Figure 1 shows composite renderings of the hot gas distribution in one of our runs, combined either with the simulated HI distribution or with the synchrotron radio emission from shocked electrons (in both cases, without observational cuts).  Clearly, while all distributions are correlated with the dark matter skeleton of the cosmic web, they all fill the cosmic web in a very different way on  $\leq 10 \rm ~Mpc$ scales, leading to relatively weak spatial cross-correlation  \citep[see e.g. our recent][]{gv20}.

As expected, the shocked gas potentially leading to radio detectable radio emission and the dense and cold gas which can promote the formation of HI sample different portions of the (mostly undetected so far) missing baryon component of the cosmic web.
As shown in the left panel of Fig. 2, where we draw contours of the portions of the gas distribution that can be detected with SKA-LOW and SKA-MID surveys \citep[][]{2019arXiv191212699B}, 
the shocked cosmic web and the HI component only overlap at high densities, while they probe entirely different gas phases elsewhere.  

For the near future,  the only observable parts of these faint components may be hunted in the neighbourhood of massive halos, and to the filaments connecting them to the rest of the cosmic web (Fig. 1). A viable observational approach will thus be to focus on the terminal parts of filaments connected to halos, possibly traced via the distribution of galaxies around them \citep[e.g.][]{2020A&A...634A..30M}.

As shown in the right panel of Fig. 2, using ROGER simulations and a new sophisticated  network reconstruction to link filaments and halos, we are monitoring the typical number of filaments connected to halos, which is an increasing function of halo mass, and only weakly dependent on the  prescriptions for gas physics ({\bf Banfi et al., submitted}\footnote{References in boldface are strictly related to the INAF-CINECA MoU.}). 

\section{Other results and future perspectives}


The topology of magnetic fields around filaments of the cosmic web has been studied by
{\bf \citet{Banfi20}}, where we quantified the impact of 
shock {\it obliquity} on the acceleration of 
cosmic ray electrons and protons (which are not accelerated by quasi-perpendicular shocks). This work highlighted that filamentary mass accretions around structures  mostly develop quasi-perpendicular shocks (due to the predominant alignment of magnetic fields with the filament outer surface), possibly explaining the low level of cosmic ray proton acceleration in galaxy clusters \citep[see also][]{wittor20}. 

The amplitude and topology of magnetic fields simulated for six different variations of input primoridal seed fields was also studied in detail in 
 {\bf \citet{va20planck}}. In this work we discussed in depth how available (or shortly incoming) day radio surveys can already be used to constrain primordial magnetic fields, significantly improving on 
present constraints based on the analysis of the Cosmic Microwave Background. 

In {\bf \citet{2020MNRAS.494.1229L}}, we simulated long light cones in order to forecast the typical maximum redshift which can be observed using the Dispersion Measure of Fast Radio Bursts, for different possible update configurations of the Northern Cross in Medicina. 

Finally, in {\bf \citet{2020arXiv200705995V}} we used the time analysis of medium sized runs to focus on the emergence of complex dynamical patterns in the evolving cosmic web, based on Information Theory \citep[following][]{2020MNRAS.491.5447V}.

In summary, all these works show that the signatures of rarefied gas in the cosmic web are weak and pose technical challenges to observations, as well as to theoretical models.
A holistic approach to model the large-scale tracers of the cosmic web will allow radio observations to be a quantitative probe of cosmic magnetism, reaching the same level of detail which is expected from incoming new radio observations, and most noticeably from the Square Kilometer Array \citep[e.g.][]{2019arXiv191212699B}. 

{\bf In closing, such systematic searches for the observable effects of variations in the physical models of cosmic gas require a secure access to HPC facilities over a multi-year time range, which would be hard to achieve without strategical MoU like the INAF-CINECA one that supported this program.}

\begin{acknowledgements}
The ENZO (enzo-project.org) simulations used for this work were produced on the Marconi-KNL Supercomputer at CINECA, under projects INA17\_C4A28 and INA17\_C5A38 and with F.V. as PI.  We also acknowledge the usage of online storage tools kindly provided by the INAF Astronomical Archive (IA2) initiative (http://www.ia2.inaf.it). F.V. and S. B. acknowledge financial support from the ERC StG MAGCOW (714196) from the H2020 initiative.
\end{acknowledgements}

\small
\bibliographystyle{aa}
\bibliography{franco3}

\begin{thebibliography}{18}
\expandafter\ifx\csname natexlab\endcsname\relax\def\natexlab#1{#1}\fi

\bibitem[{{Banfi} {et~al.}(2020){Banfi}, {Vazza}, \& {Wittor}}]{Banfi20}
{Banfi}, S., {Vazza}, F., \& {Wittor}, D. 2020, arXiv e-prints,
  arXiv:2006.10063

\bibitem[{{Braun} {et~al.}(2019){Braun}, {Bonaldi}, {Bourke}, {Keane}, \&
  {Wagg}}]{2019arXiv191212699B}
{Braun}, R., {Bonaldi}, A., {Bourke}, T., {Keane}, E., \& {Wagg}, J. 2019,
  arXiv e-prints, arXiv:1912.12699

\bibitem[{{Bryan} {et~al.}(2014){Bryan}, {Norman}, {O'Shea}, {Abel}, {Wise},
  {Turk}, {Reynolds}, {Collins}, {Wang}, {Skillman}, {Smith}, {Harkness},
  {Bordner}, {Kim}, {Kuhlen}, {Xu}, {Goldbaum}, {Hummels}, {Kritsuk}, {Tasker},
  {Skory}, {Simpson}, {Hahn}, {Oishi}, {So}, {Zhao}, {Cen}, {Li}, \& {Enzo
  Collaboration}}]{enzo14}
{Bryan}, G.~L., {Norman}, M.~L., {O'Shea}, B.~W., {et~al.} 2014, \apjs, 211, 19

\bibitem[{{Gheller} \& {Vazza}(2020)}]{gv20}
{Gheller}, C. \& {Vazza}, F. 2020, \mnras, 494, 5603

\bibitem[{{Hoeft} \& {Br{\"u}ggen}(2007)}]{hb07}
{Hoeft}, M. \& {Br{\"u}ggen}, M. 2007, \mnras, 375, 77

\bibitem[{{Horii} {et~al.}(2017){Horii}, {Asaba}, {Hasegawa}, \&
  {Tashiro}}]{2017PASJ...69...73H}
{Horii}, T., {Asaba}, S., {Hasegawa}, K., \& {Tashiro}, H. 2017, \pasj, 69, 73

\bibitem[{{Locatelli} {et~al.}(2020){Locatelli}, {Bernardi}, {Bianchi},
  {Chiello}, {Magro}, {Naldi}, {Pilia}, {Pupillo}, {Ridolfi}, {Setti}, \&
  {Vazza}}]{2020MNRAS.494.1229L}
{Locatelli}, N.~T., {Bernardi}, G., {Bianchi}, G., {et~al.} 2020, \mnras, 494,
  1229

\bibitem[{{Malavasi} {et~al.}(2020){Malavasi}, {Aghanim}, {Tanimura},
  {Bonjean}, \& {Douspis}}]{2020A&A...634A..30M}
{Malavasi}, N., {Aghanim}, N., {Tanimura}, H., {Bonjean}, V., \& {Douspis}, M.
  2020, \aap, 634, A30

\bibitem[{{Nicastro}(2016)}]{2016xnnd.confE..27N}
{Nicastro}, F. 2016, in XMM-Newton: The Next Decade, 27

\bibitem[{{Popping} {et~al.}(2015){Popping}, {Meyer}, {Staveley-Smith},
  {Obreschkow}, {Jozsa}, \& {Pisano}}]{2015arXiv150101077P}
{Popping}, A., {Meyer}, M., {Staveley-Smith}, L., {et~al.} 2015, ArXiv e-prints

\bibitem[{{Subramanian}(2016)}]{sub16}
{Subramanian}, K. 2016, Reports on Progress in Physics, 79, 076901

\bibitem[{{Tramonte} {et~al.}(2019){Tramonte}, {Ma}, {Li}, \&
  {Staveley-Smith}}]{2019MNRAS.489..385T}
{Tramonte}, D., {Ma}, Y.-Z., {Li}, Y.-C., \& {Staveley-Smith}, L. 2019, \mnras,
  489, 385

\bibitem[{{Vazza}(2020{\natexlab{a}})}]{2020MNRAS.491.5447V}
{Vazza}, F. 2020{\natexlab{a}}, \mnras, 491, 5447

\bibitem[{{Vazza}(2020{\natexlab{b}})}]{2020arXiv200705995V}
---. 2020{\natexlab{b}}, arXiv e-prints, arXiv:2007.05995

\bibitem[{Vazza {et~al.}(2017)Vazza, Brueggen, Gheller, Hackstein, Wittor, \&
  Hinz}]{va17cqg}
Vazza, F., Brueggen, M., Gheller, C., {et~al.} 2017, Classical and Quantum
  Gravity

\bibitem[{{Vazza} {et~al.}(2020){Vazza}, {Paoletti}, {Banfi}, {Finelli},
  {Gheller}, {O'Sullivan}, \& {Br{\"u}ggen}}]{va20planck}
{Vazza}, F., {Paoletti}, D., {Banfi}, S., {et~al.} 2020, arXiv e-prints,
  arXiv:2009.01539

\bibitem[{{Wang} {et~al.}(2010){Wang}, {Abel}, \& {Kaehler}}]{wang10}
{Wang}, P., {Abel}, T., \& {Kaehler}, R. 2010, \na, 15, 581

\bibitem[{{Wittor} {et~al.}(2020){Wittor}, {Vazza}, {Ryu}, \&
  {Kang}}]{wittor20}
{Wittor}, D., {Vazza}, F., {Ryu}, D., \& {Kang}, H. 2020, \mnras, 495, L112

\end{thebibliography}

\end{document}